\documentclass[twocolumn,amsmath,amssymb]{revtex4}

\usepackage{graphicx}
\usepackage{dcolumn}
\usepackage{bm}

\begin{document}


\title{Tail-scope: Using friends to estimate heavy tails of degree distributions \\
in large-scale complex networks}

\author{Young-Ho Eom}
\affiliation{Laboratoire de Physique Th\'eorique du CNRS, IRSAMC,
Universit\'e de Toulouse, UPS, F-31062 Toulouse, France}
\affiliation{IMT Institute for Advanced Studies Lucca, Piazza San
Francesco 19, Lucca 55100, Italy}
\author{Hang-Hyun Jo}
\affiliation{BK21plus Physics Division and Department of Physics, Pohang University of Science and Technology, Pohang 790-784, Republic of Korea} \affiliation{BECS, Aalto University School of Science, P.O. Box 12200, FI-00076, Finland}


\begin{abstract}
Many complex networks in natural and social phenomena have often been characterized by heavy-tailed degree distributions. However, due to rapidly growing size of network data and concerns on privacy issues about using these data, it becomes more difficult to analyze complete data sets. Thus, it is crucial to devise effective and efficient estimation methods for heavy tails of degree distributions in large-scale networks only using local information of a small fraction of sampled nodes. Here we propose a \emph{tail-scope} method based on local observational bias of the friendship paradox. We show that the tail-scope method outperforms the uniform node sampling for estimating heavy tails of degree distributions, while the opposite tendency is observed in the range of small degrees. In order to take advantages of both sampling methods, we devise the hybrid method that successfully recovers the whole range of degree distributions. Our tail-scope method shows how structural heterogeneities of large-scale complex networks can be used to effectively reveal the network structure only with limited local information.
\end{abstract}

\maketitle

\section*{Introduction}

Complex networks have served as a powerful mathematical framework
to describe complex systems of nature, society, and
technology~\cite{Boccaletti2006,Newman2010,Barabasi2004,Lazer2009,Vespignani2011}.
Most complex networks obtained from complex systems are known to
be heterogeneous in various
aspects~\cite{Barabasi1998,Watts1998,Newman2002,Fortunato2010}.
One of distinctive heterogeneous features in complex networks is
the heavy-tailed degree distribution: A small number of highly
connected nodes coexist with the large number of lowly connected
nodes. Highly connected nodes or hubs found in heavy tails have
significant roles on the evolution of complex networks and
dynamics on such networks. For examples, the existence of hubs
leads networks to endemic states in epidemic
spreading~\cite{Pastor-Satorras2001,Castellano2012}, makes
networks vulnerable to intended attacks~\cite{Albert2000}, and
contributes to the key functions of biological
systems~\cite{Jeong2001,Han2004,Zotenko2008}. Therefore,
identifying the degree distribution and particularly hubs in the
heavy tail of degree distribution is the essential step for the
network analysis~\cite{Clauset2009}.

Owing to the rapid development of digital technologies,
a huge amount of network data is being generated and recorded. In
particular, the network data from social media like Twitter and
Wikipedia contain tens of millions to billion nodes (users or
articles). The role of social media on social dynamics such as
public opinion formation, information diffusion, and
popularity~\cite{Centola2010,Bakshy2012,Christakis2007} is getting
more crucial, requiring us to timely monitor the large-scale
dynamics and to identify the network structure underlying these
dynamics~\cite{Lazer2009,Castello2009}. However, since the social
media are constantly growing and changing, the acquisition and
analysis of complete network data is an extremely tricky task.
Further, increasing public concerns on privacy issues about using
these data can inhibit us from analyzing the complete network
data~\cite{Garcia-Herranz2014}.

Because of the above difficulties, degree distributions of complex
networks need to be estimated based on partial information or by
sampling nodes from networks. The simplest method could be to
sample nodes randomly, which is called uniform node sampling
(UNS). Since the number of nodes corresponding to the tail part of
distribution is typically very small, those nodes are rarely
sampled, limiting the \emph{sampling resolution}.
Accordingly, much larger statistical fluctuations are expected for
the tail part of degree distribution estimated by UNS, when
compared to its body part.

The friendship paradox (FP)~\cite{Feld1991,Hodas2013,Eom2014GFP}
can shed light on how to effectively estimate the heavy tails of
degree distributions. The FP states that the degree of an
individual is on average smaller than the average degree of its
friends or neighbors. The underlying mechanism behind the FP is
the observational bias such that highly connected nodes are more
likely to be observed by their neighbors. One can take advantage
of this observational bias for the effective sampling of highly
connected nodes. Indeed, the group made of friends of randomly
chosen nodes turns out to contain highly connected nodes more than
the group made of uniformly sampled
nodes~\cite{Eom2014GFP,Avrachenkov2014}. Further, the FP has also
been used for early detection of contagious
outbreaks~\cite{Garcia-Herranz2014,Christakis2010} and natural
disaster~\cite{Kryvasheyeu2014}, and for designing efficient
immunization strategy~\cite{Cohen2003}. These are mainly based on
the observation of activities of highly connected nodes via the FP
rather than uniformly sampled nodes.

In this paper, we devise a novel sampling method, called
tail-scope, to effectively estimate the heavy tails of degree
distributions in large-scale complex networks.
We exploit the observational bias of FP as a magnifying glass to
observe heavy tails with better resolution and to overcome the
resolution limit in the UNS. It is shown that the tail-scope
method estimates heavy tails of empirical degree distributions in
large-scale networks more accurately than the UNS. Finally, we
suggest a hybrid sampling method taking advantages of both UNS and
tail-scope methods to recover the whole range of degree
distribution.

\section*{Results}

\subsection*{Tail-scope: Estimating the tail of degree distribution using the friendship paradox}

We consider a directed network $G=G(N,L)$ with $N$ nodes and $L$
directed links. In case of undirected networks, each undirected
link is considered as two directed links in both directions. For a
node $i$, the in-degree $k_i$ represents the number of incoming
links to $i$ from $i$'s in-neighbors, and the in-degree
distribution is denoted by $P(k)$. Similarly, one can define the
out-degree as the number of out-neighbors.

Our goal is to effectively estimate the heavy tail of in-degree
distribution, i.e., the region of $k\gg 1$, by using partial
information such as by sampling $n$ nodes with $n\ll N$. The
observational bias of friendship paradox (FP) indicates that
observation via friends can lead to the larger number of high
degree nodes than that by the uniform node sampling (UNS), because
the chance of a node being observed by its neighbors is
proportional to the degree of the node. For this, we randomly
choose $n$ directed links and construct a set of nodes reached by
following those links. The probability of finding a node of
in-degree $k$ in the set is proportional to $kP(k)$ not to $P(k)$,
which we denote by $\tilde Q(k)$:
\begin{equation}
\tilde Q(k) \propto kP(k).
\label{eq:1}
\end{equation}
Then we obtain the estimated in-degree distribution as
\begin{equation}
\tilde P(k) \propto \frac{\tilde Q(k)}{k}.
\label{eq:2}
\end{equation}
Thanks to the observational bias of FP, the estimated $\tilde
P(k)$ has the larger number of highly connected nodes and hence
less statistical fluctuation for the tail part than when the UNS
is used. Our method can be called \emph{tail-scope}. Precisely,
the sampling resolution characterized by the cutoff $k_c(n)$ of
the distribution is higher for the tail-scope method than for the
UNS.

In order to demonstrate the effectiveness of tail-scope method for
estimating the heavy tail of the distribution, we consider a
network showing the power-law in-degree distribution with
power-law exponent $\alpha>2$ and minimum in-degree $k_{\rm min}$:
\begin{equation}
  P(k)= (\alpha-1)k_{\rm min}^{\alpha-1}k^{-\alpha},
\end{equation}
where we have assumed for convenience that the in-degree $k$ is a
continuous variable. At first, by randomly choosing $n$ nodes
(i.e., by UNS) we obtain the estimated in-degree distribution
$P_{_{\rm UN}}(k)$ that is expected to be $\propto k^{-\alpha}$.
Due to the finiteness of $n$, we find the natural cutoff to the
power-law tail as
\begin{equation}
  P_{_{\rm UN}}(k)\propto k^{-\alpha}e^{-k/k_c},
\end{equation}
where $k_c$ can be characterized by the condition
\begin{equation}
  \frac{1}{n}=\int_{k_c}^\infty P(k)dk,
\end{equation}
leading to
\begin{equation}
  k_c=k_{\rm min} n^{1/(\alpha-1)}.
\end{equation}
Next, for the tail-scope method, we expect from $\tilde Q(k)\propto kP(k)$ that
\begin{eqnarray}
  \tilde Q(k)&\propto& k^{-(\alpha-1)}e^{-k/k'_c},\\
  k'_c&=&k_{\rm min} n^{1/(\alpha-2)}.
\end{eqnarray}
Then one gets the estimated in-degree distribution in Eq.~(\ref{eq:2}):
\begin{equation}
\tilde P(k) \propto k^{-\alpha}e^{-k/k'_c}.
\end{equation}
It is evident that the sampling resolution $k'_c$ for the tail-scope case is higher than $k_c$ for the UNS, precisely,
\begin{equation}
  \frac{k'_c}{k_c}=n^{1/[(\alpha-2)(\alpha-1)]}>1.
\end{equation}
Therefore, our tail-scope method indeed outperforms the UNS for estimating the tail of the distribution. Since the tail-scope method is based on the uniform link sampling, it can also be called \emph{link tail-scope}, mainly in order to distinguish from \emph{node tail-scope} to be discussed in the next Subsection.

We numerically test our calculations by constructing the Barab\'asi-Albert (BA) scale-free network~\cite{Barabasi1998} with $N=10^6$, $k_{\rm min}=2$, and $\alpha=3$, and then by sampling $n=500$ nodes. From the calculations, we expect that $k_c\approx 45$ and $k'_c\approx 1000$, which are numerically confirmed as shown in Fig~\ref{fig:1}(A). In the figures, we have used the complementary cumulative distribution function (CCDF), defined as $F(k)=\sum_{k'=k}^\infty P(k')$, for clearer visualization.

\subsection*{Node-based tail-scope method}

Our tail-scope method is based on the uniform link sampling.
However, in many realistic situations, we can use only the
node-based sampling not the link-based sampling. For instance,
most application programming interfaces (APIs) of social media
like Twitter allow us to retrieve only the user-specific
information rather than the relationship-based ones. Thus it is
necessary to develop a sampling method using node-based data but
aimed to simulate the link tail-scope method.

As social media APIs allow to get only user-specific local
information in most cases, we assume that whenever a node is
sampled or retrieved, we get the set of in- and out-neighbors for
the sampled node. These constraints inevitably introduce
correlations between sampled links, implying that any node-based
tail-scope methods cannot be exactly mapped to the link tail-scope
method. In addition, we assume that the number of retrievals,
i.e., sampling size, is strictly limited to $n$ for the fair
comparison to
other sampling methods, e.g., the UNS. We propose the node tail-scope method as follows.\\

\emph{Node tail-scope method:}
\begin{itemize}
  \item Step 1. Randomly choose $n/2$ nodes (called primary nodes) from the network and retrieve their out-neighbors to construct a set $A$ of those out-neighbors.
  \item Step 2. Randomly choose $n/2$ nodes from the set $A$ and retrieve their in-degrees to construct the distribution $Q_{_{\rm NT}}(k)$.
  \item Step 3. Obtain the estimated in-degree distribution $P_{_{\rm NT}}(k)$ from $Q_{_{\rm NT}}(k)/k$.\\
\end{itemize}

Here the subscript NT of distributions is the abbreviation of node tail-scope. Note that as the total number of retrievals is limited to $n$, we use $n/2$ retrievals for getting out-neighbors, and the rest $n/2$ retrievals for getting in-degrees. However, there are more high degree nodes sampled than when the UNS is used, leading to the higher resolution for the tail-scope method. For a node sampled several times in Step 2, we consider each sampling as a different case.

\begin{figure*}[ht]
\begin{center}
\includegraphics[width=60mm,angle=-90]{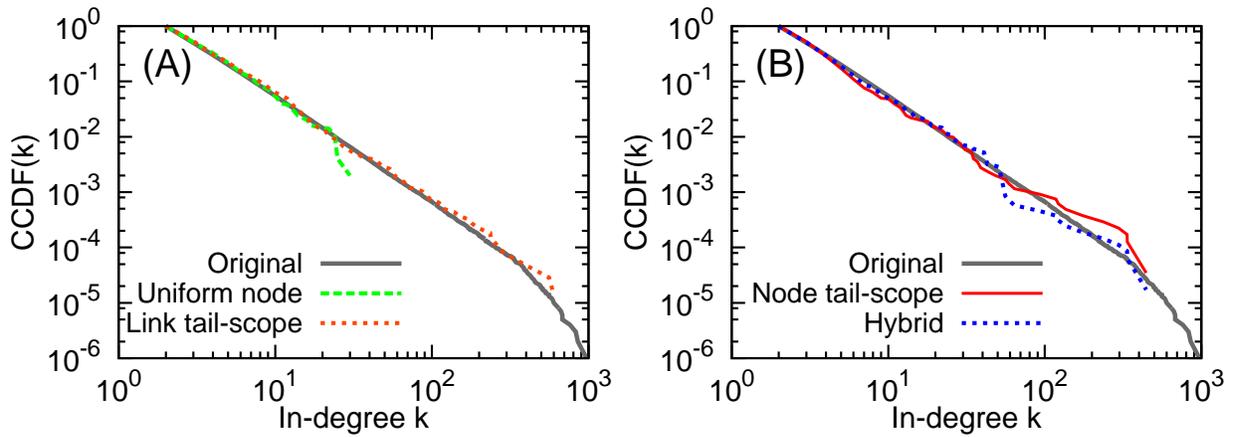}
  \caption{Comparison of in-degree distributions estimated by uniform node sampling, link tail-scope, node tail-scope, and hybrid methods to the original distribution for the Barab\'asi-Albert scale-free network with $N=10^6$ and minimum in-degree $k_{\rm min}=2$. The sample size is $n=500$. In all cases, complementary cumulative distribution functions (CCDFs) are presented.}
\label{fig:1}
\end{center}
\end{figure*}

By using the same BA network in the previous Subsection, we
compare the performance of node tail-scope, shown in
Fig.~\ref{fig:1}(B) to that of link tail-scope in
Fig.~\ref{fig:1}(A). It is observed that there is no significant
difference between two results.

\subsection*{Performance of the node tail-scope method}

In order to empirically compare the performance of node tail-scope method to the UNS, we consider several large-scale complex networks: four undirected networks and four directed networks. For details of these networks, see the Method Section and Table~\ref{table:NetStat}. From now on, we use the sample size $n=1000$ in all cases. As mentioned, such small number of $n$ is due to the practical constraint on the number of retrievals. When the constraint is relaxed, other sampling methods using graph traversal techniques (e.g., breadth first search) can be used, inducing more complicated observational biases~\cite{Kurant2011}.

\begin{figure*}[ht]
\begin{center}
\includegraphics[width=136mm,angle=-90]{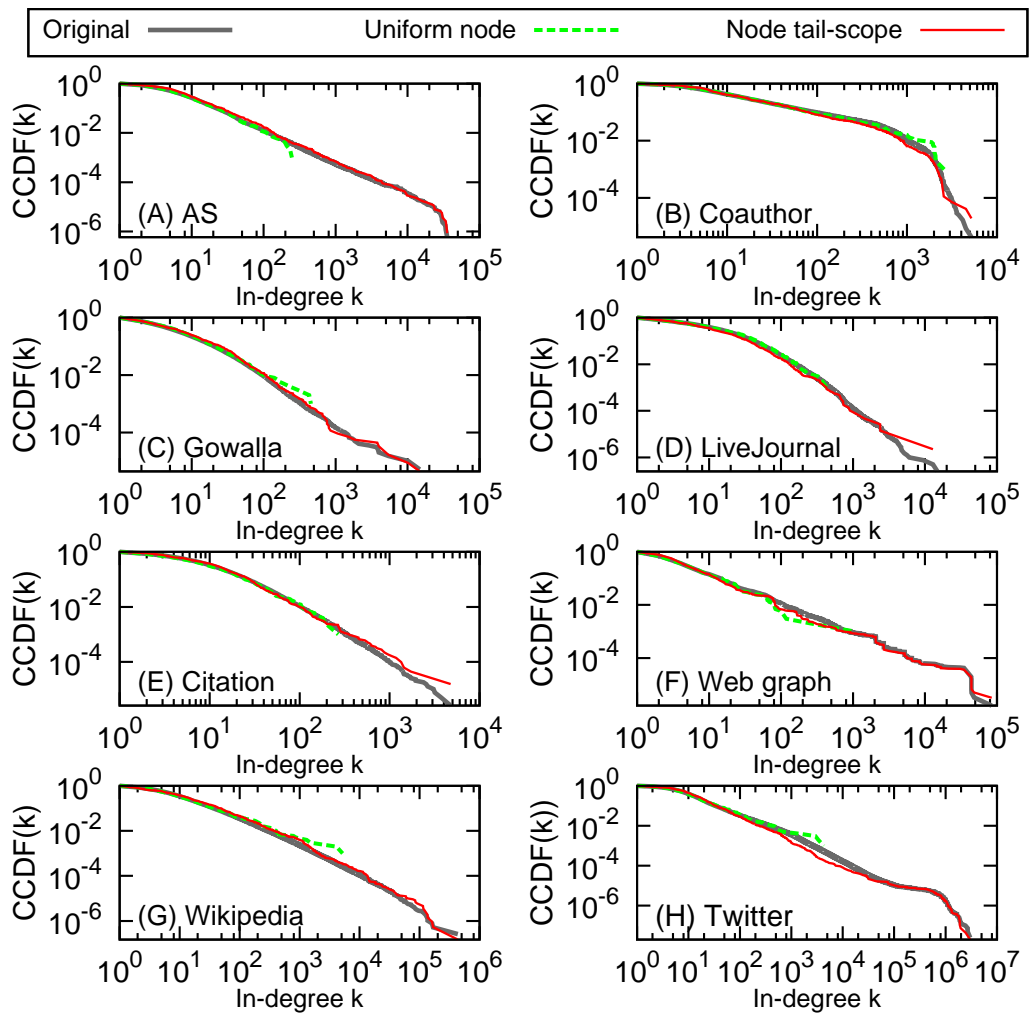}
  \caption{Comparison of in-degree distributions estimated by uniform node sampling and node tail-scope methods to the original distributions for several empirical directed and undirected networks. The sample size is $n=1000$. In all cases, complementary cumulative distribution functions (CCDFs) are presented. For the details of the networks, see the Method Section and Table~\ref{table:NetStat}.}
  \label{fig:tailscope}
\end{center}
\end{figure*}

Figure~\ref{fig:tailscope} shows estimated in-degree distributions $P_{_{\rm NT}}(k)$ (node tail-scope) and $P_{_{\rm UN}}(k)$ (UNS), in comparison to the original in-degree distribution $P(k)$ obtained from the complete set of nodes in the network. The agreements between original distributions and the distributions by node tail-scope method in the tail parts are remarkable, while some fluctuations are observed in the body parts. On the other hand, the distributions by the UNS show good agreements with the original distributions in the body parts, not in the tail parts. Note that the sample size $n=1000$ is much smaller than the network size $N$ ranging from hundreds of thousands to tens of millions nodes (see Table~\ref{table:NetStat}). We find that the results using $n=2000$ and $n=4000$ are qualitatively the same as the case of $n=1000$.

For the quantitative comparison of performance by different sampling methods, we use Kolmogorov-Smirnov (KS) static $D$, defined as the maximum difference between two CCDFs. The KS $D$-static is mainly used as a part of KS test to reject null hypothesis. For example, it has been used to test if a given distribution has a power-law tail~\cite{Clauset2009}. In this paper, we simply use $D$-static to measure the agreement between the original in-degree distribution and the estimated in-degree distribution by each sampling method. The $D$-static for the node tail-scope method is obtained as
\begin{equation}
  D_{_{\rm NT}} = \max_k |F_{_{\rm NT}}(k)-F(k)|,
\end{equation}
where $F(k)$ denotes the CCDF of the original in-degree distribution, and $F_{_{\rm NT}}(k)$ denotes the CCDF of $P_{_{\rm NT}}(k)$. Similarly, $D_{_{\rm UN}}$ is defined for the UNS. The smaller $D$-static implies the better agreement to the original distribution.

\begin{figure*}[ht]
\begin{center}
\includegraphics[width=60mm,angle=-90]{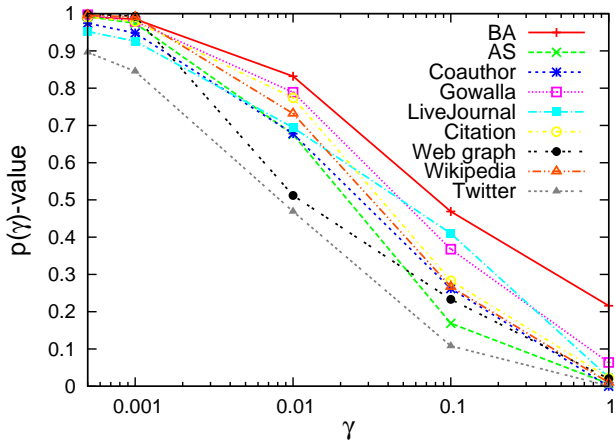}
  \caption{Performance of the node tail-scope method compared to the uniform node sampling for all the considered networks. $p(\gamma)$ is calculated by Eq.~(\ref{eq:pvalue}) but with Kolmogorov-Smirnov $D$-statics defined only for the range of $k\geq k_0$, where $\gamma=F(k_0)$. The smaller $\gamma$ corresponds to the larger $k_0$. The larger $p(\gamma)$-values imply the better performance of the node tail-scope method than the uniform node sampling. To get $p$-values, we used $1000$ realizations of sampling, for each of which the sample size is $n=1000$.}
\label{fig:Pvaluetailscope}
\end{center}
\end{figure*}

Then, we define a $p$-value to compare the two considered sampling methods. The $p$-value represents the probability that the distribution by node tail-scope method has the smaller $D$-static with the original distribution than the distribution by the UNS, i.e.,
\begin{equation}
  p=\Pr( D_{_{\rm NT}}<D_{_{\rm UN}}).
  \label{eq:pvalue}
\end{equation}
To focus on the tail part of the distribution, we compare the CCDFs only for the region of $k\geq k_0$, or equivalently for the fraction $\gamma$ of high degree nodes, where $\gamma=F(k_0)$. The case of $\gamma=1$ corresponds to the comparison for the entire range of in-degree. Figure~\ref{fig:Pvaluetailscope} shows the values of $p(\gamma)$ for different ranges of in-degree and for each considered network. It is found for all networks that the node tail-scope method clearly outperforms the UNS for the tail parts. The opposite tendency is observed when the entire range of the distribution is compared, because the UNS outperforms the node tail-scope for estimating the body part of the distribution. Since the sample size $n$ is limited, the larger number of high degree nodes for the node tail-scope method results in the smaller number of low degree nodes and hence the larger fluctuations than the case of UNS.


As mentioned, since the node tail-scope method inevitably
introduces correlations between sampled links, we now consider
possible effects of degree correlations on the performance of node
tail-scope method. As shown in Fig.~\ref{fig:1}, in the case of BA
scale-free network with negligible degree correlation, the
performance difference between the link tail-scope and the node
tail-scope methods is not significant. We draw the same conclusion
for considered empirical networks showing degree correlations, in
terms of non-zero assortativity coefficients~\cite{Newman2002}.
For example, the assortativity coefficients are $r\approx -0.08$
(AS), $-0.029$ (Gowalla), $0.467$ (Coauthorship), and $0.045$
(LiveJournal). These observations support the validity of our
methods.

\begin{figure*}[ht]
\begin{center}
\includegraphics[width=60mm,angle=-90]{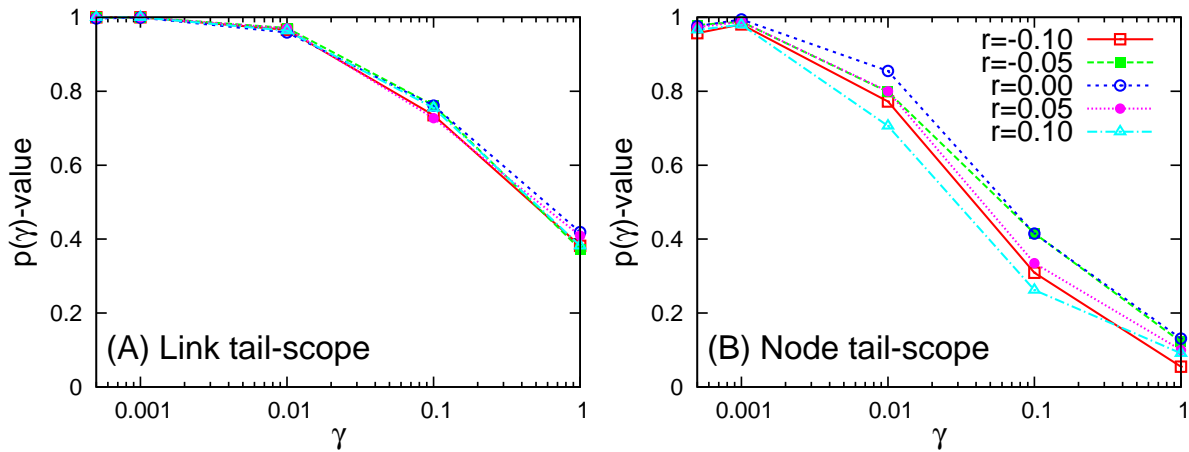}
  \caption{Performance of (A) link tail-scope method and of (B) node tail-scope method compared to the uniform node sampling for correlated scale-free networks with $N=50000$ and degree exponent $\alpha=2.7$ for $r=-0.1, -0.05, 0, 0.05, 0.1$, where $r$ denotes the assortativity coefficient~\cite{Newman2002}. $p(\gamma)$ is calculated by Eq.~(\ref{eq:pvalue}) but with Kolmogorov-Smirnov $D$-statics defined only for the range of $k\geq k_0$, where $\gamma=F(k_0)$. The smaller $\gamma$ corresponds to the larger $k_0$. The larger $p(\gamma)$-values imply the better performance of the link (node) tail-scope method than the uniform node sampling. To get $p$-values, we used $1000$ realizations of sampling, for each of which the sample size is $n=500$.}
\label{fig:PvalueCorrelated}
\end{center}
\end{figure*}

For making sure the validity of our methods for networks with non-zero degree correlation, we numerically consider correlated scale-free networks with tunable degree correlation used in~\cite{Jo2014}. By using several scale-free networks with $N=50000$, degree exponent $2.7$ for $r=-0.1, -0.05, 0, 0.05, 0.1$, we obtain the $p$-values for each case. As expected, the link tail-scope method is barely influenced by the correlation (Fig.~\ref{fig:PvalueCorrelated}(A)). The node tail-scope method shows some effects of correlation but still gives us better sampling results than when UNS is used (Fig.~\ref{fig:PvalueCorrelated} (B)). Overall, the sampling results can be affected if the degree correlation is quite strong. However, our method still performs better for sampling the tail parts than the UNS.

\subsection*{Hybrid method for recovering the whole distribution}

It is evident that the UNS and the node tail-scope method are good at sampling low and high degree nodes, respectively. In order to take advantages of both methods, we suggest the \emph{hybrid method} for recovering the whole range of the distribution. It is notable that at Step $1$ in our node tail-scope method, $n/2$ primary nodes are randomly chosen and hence their in-degrees can be utilized for the low degree region. From the primary nodes, we get the in-degree distribution $P_{_{\rm NT0}}(k)$. Then the hybrid distribution is obtained by
\begin{equation}
  P_{_{H}}(k) = a P_{_{\rm NT0}}(k) +(1-a) P_{_{\rm NT}}(k).
  \label{eq:hybrid}
\end{equation}
The weight parameter $a\in [0,1]$ can be chosen according to which part of the distribution is focused. Here we set as $a=0.5$.

\begin{figure*}[ht]
\begin{center}
\includegraphics[width=66mm,angle=-90]{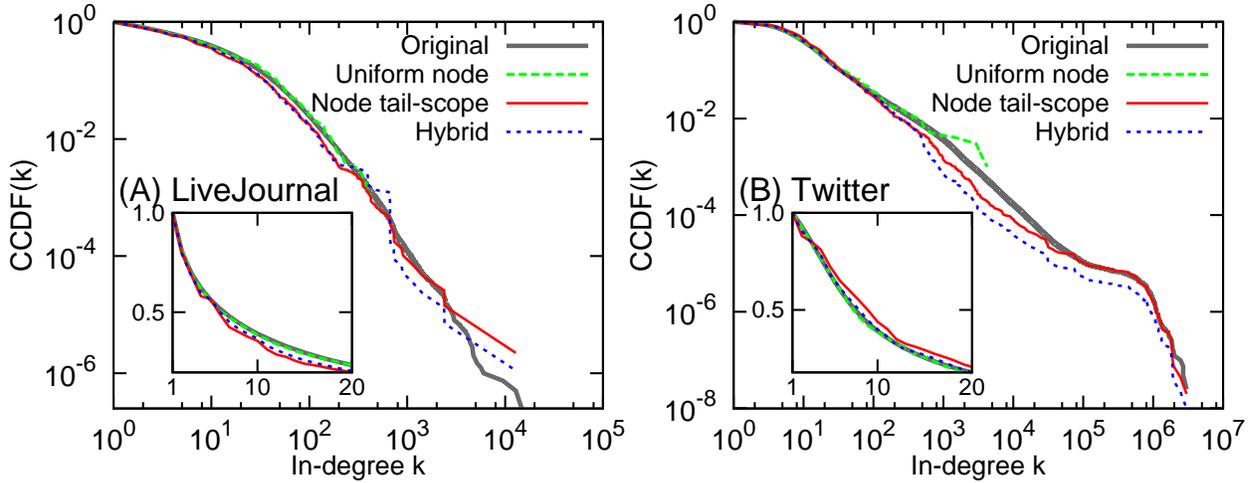}
  \caption{Comparison of in-degree distributions estimated by uniform node sampling, node tail-scope, and hybrid methods to the original distributions for networks of LiveJournal (A) and Twitter (B). The insets show results for the range of $k\leq 20$. The sample size is $n=1000$. In all cases, complementary cumulative distribution functions (CCDFs) are presented.}
\label{fig:hybrid}
\end{center}
\end{figure*}

\begin{figure*}[ht]
\begin{center}
\includegraphics[width=60mm,angle=-90]{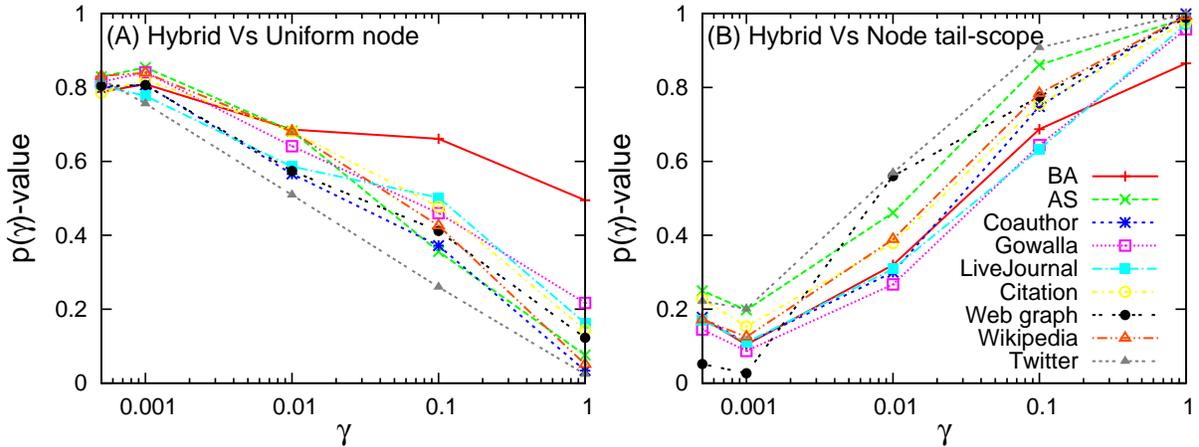}
  \caption{Performance of the hybrid method compared to the uniform node sampling (A) and to the node tail-scope method (B) for all the considered networks. $p(\gamma)$ is calculated by Eq.~(\ref{eq:pvalue}) but with Kolmogorov-Smirnov $D$-statics defined only for the range of $k\geq k_0$, where $\gamma=F(k_0)$. The smaller $\gamma$ corresponds to the larger $k_0$. The larger $p(\gamma)$-values imply the better performance of the hybrid method than the uniform node sampling (A) or the node tail-scope method (B). To get $p$-values, we used $1000$ realizations of sampling, for each of which the sample size is $n=1000$.}
\label{fig:Pvaluehybrid}
\end{center}
\end{figure*}

The hybrid method performs well for the BA network in Fig.~\ref{fig:1}(B) as well as for empirical networks, two of which are shown in Fig.~\ref{fig:hybrid}. As expected, the distributions estimated by the hybrid method fit the original distributions better than the UNS for the tail parts, and better than the node tail-scope method for the body parts (see insets in Fig.~\ref{fig:hybrid}). These findings are also consistent with the values of $p(\gamma)$ shown in Fig.~\ref{fig:Pvaluehybrid}: The larger values of $p(\gamma)$ for small values of $\gamma$ in Fig.~\ref{fig:Pvaluehybrid}(A) imply the better performance of the hybrid method than the UNS for the tail parts. The larger values of $p(\gamma)$ for large values of $\gamma$ in Fig.~\ref{fig:Pvaluehybrid}(B) imply the better performance of the hybrid method than the node tail-scope for the body parts. Therefore, we conclude that the hybrid method successfully recovers the whole range of in-degree distributions, by taking advantages of both the UNS and the node tail-scope methods. Other values of $a=0.25$ and $a=0.75$ have been also tested and all results are as expected.

\section*{Discussion}

Modern societies have been shaped by large-scale networked systems
like World Wide Web, social media, and transportation systems.
Monitoring global activities and identifying the network structure
of these systems are of utmost importance in better understanding
collective social dynamics. However, increasing size of data from
these systems and growing concerns on privacy issues about using
these data make the exhausted analysis of complete data sets
infeasible. Thus, effective and efficient estimation of large-scale networks
based on the small sample size or partial information is
necessary. One of the simplest method could be uniform node
sampling (UNS). The UNS has drawbacks in particular for estimating
the heavy tails of degree distributions, due to the limited
sampling resolution and large statistical fluctuations. Since high
degree nodes found in the heavy tails are in many cases very
important to characterize the structure and dynamics of complex
networks, we propose the tail-scope method, which is the effective and efficient
sampling method for estimation of heavy tails of
degree distributions.

Provided that the sample size is limited, it is inevitable that
the larger number of high degree nodes by the tail-scope method
leads to the smaller number of low degree nodes than when the UNS
is used. In order to take advantages of both the tail-scope and
the UNS, we propose the hybrid method to recover the whole range
of degree distributions. In this paper, we have considered a very
simple form of hybrid method by superposing the estimated degree
distributions of the UNS and the tail-scope. It turns out that the
hybrid method performs better than the UNS for the tail parts, and
better than the tail-scope for the body parts. Devising more
general and better hybrid methods will be interesting as a future
work, e.g., one can use the degree-dependent weight parameter $a$
in Eq.~(\ref{eq:hybrid}).

Our tail-scope method can be also used for estimating high
attribute nodes found in the heavy tail of attribute distribution.
The attribute of a node can be its activity, income, happiness,
and so on. Recently, the generalized friendship paradox (GFP) has
been observed and analyzed in complex
networks~\cite{Eom2014GFP,Jo2014}. The GFP states that the
attribute of a node is on average lower than the average attribute
of its neighbors. In the network showing the positive correlation
between degrees and attributes, high degree nodes tend to have
higher attributes. It implies that the high attribute nodes are
more likely to be observed by their neighbors. Such generalized
observational bias can be exploited to effectively estimate high
attribute nodes who play important roles, e.g., in early detection
of new trends or in designing efficient immunization strategies.
Thus, it would be very interesting to generalize our tail-scope
method to other attributes of nodes, especially for the
large-scale complex networks.

Our tail-scope method shows how structural heterogeneities can
help us reveal the network structure only with limited
information. By exploiting such heterogeneities of complex
networks we can properly evaluate priority and importance of each
node in the networks. It is getting more important to better
understand the heterogeneities since they are key features
characterizing the complexity of large-scale networks.

\section*{Methods}

\subsection*{Data description}

In this paper, we consider eight empirical networks: four of them
are undirected and the others are directed. The summary of the
networks is presented in Table~\ref{table:NetStat}. The detailed
feature of each network is as following.

\emph{AS.} We used an Autonomous Systems (ASs) data set on
Internet topology graph constructed in~\cite{Leskovec2005}. The
nodes are autonomous systems and the links are formed where two
ASs exchange traffic flows. The size of network is $N=1696415$.

\emph{Coauthorship.} We used a coauthorship network constructed in~\cite{Eom2014GFP}. The nodes are scientists and the links are formed whenever two scientists coauthored the paper. The network size is $N=242592$.

\emph{Gowalla.} We used a Gowalla friendship network constructed in~\cite{Cho2011}. Gowalla is a location-based social networking service. Each user defines a node. The network size is $N=196562$.

\emph{LiveJournal.} We used a LiveJournal friendship network constructed in~\cite{Yang2012}. Livejournal.com is a social networking service for blog, journal, and diary. The nodes are users of LiveJournal and the users can declare friendship to another user, defining a link. The network size is $N=3997962$.

\emph{Citation.} We used a citation network constructed in~\cite{Eom2011}. The network is based on the bibliographic database from 1893 to 2009 provided by American Physical Society (APS). The nodes are articles published in APS journal such as Physical Review Letters or Physical Review E and the directed links represent the citation relation between articles. The network size is $N=463349$.

\emph{Web graph.} We used a web graph constructed in~\cite{Leskovec2009}. The nodes represent webpages in the domains of berkely.edu and stanford.edu domains, and the links are hyperlink between webpages. The network size is $N=685230$.

\emph{Wikipedia.} We used an English Wikipedia network constructed in~\cite{Eom2014Wiki}. The Wikipedia data set was collected in February 2013. The nodes are English Wikipedia articles and the links are hyperlinks between those articles. The network size is $N=4212493$.

\emph{Twitter.} We used a Twitter users network constructed in~\cite{Kwak2010}. The nodes are Twitter users and the links between users represent the following relations in Twitter. The network size is $N=41652230$.

\section*{Acknowledgements}

The authors thank American Physical Society for providing Physical Review bibliographic data. Y.-H.E. acknowledges support from the EC FET Open project ``New tools and algorithms for directed network analysis'' (NADINE number 288956) and the EC FET project ``Financial systems simulation and policy modelling (SIMPOL)'' - No. 610704. H.-H.J. acknowledges financial support by Aalto University postdoctoral program and by Mid-career Researcher Program through the National Research Foundation of Korea (NRF) grant funded by the Ministry of Science, ICT and Future Planning (2014030018), and Basic Science Research Program through the National Research Foundation of Korea (NRF) grant funded by the Ministry of Science, ICT and Future Planning (2014046922).

\section*{Author Contribution}

Y.-H.E. and H.-H.J. designed research, wrote, reviewed, and
approved the manuscript. Y.-H.E. devised the algorithm and
performed data collection and analysis.

\section*{Additional Information}
Competing financial interests: The authors declare no competing
financial interests.
{}


\begin{table*}[!ht]
  \caption{Basic statistics of empirical undirected and directed networks. $N$ denotes the total number of nodes and $\langle k \rangle$ denotes the average in-degree. The isolated nodes have been excluded for the analysis.}
\label{table:NetStat}
\begin{center}
\begin{tabular}{c|c|c|c|c|c}
 \hline
Undirected network & $N$ & $\langle k \rangle $ & Directed network & $N$ & $\langle k \rangle $ \\
 \hline
AS & 1696415 & 13.1 & Citation & 463349 & 12.2  \\
Coauthorship & 242592 & 59.6 & Web graph & 685230 & 12.3 \\
Gowalla & 196562 & 9.7  & Wikipedia & 4212493 & 26.4 \\
LiveJournal & 3997962 & 17.3 & Twitter & 41652230 & 36.6 \\
 \hline
\end{tabular}
\end{center}
\end{table*}

\end{document}